\documentclass[preprint,aps,prd,nofootinbib,floatfix,amsmath,amssymb]{revtex4-2}
\usepackage{graphicx}
\usepackage[utf8x]{inputenc}
\usepackage{color}
\usepackage{subfig}
\usepackage{multirow}
\usepackage{slashed}
\usepackage{cancel}
\usepackage{hyperref}
\usepackage{amsmath}
\usepackage{pdfsync}

\begin{document}

\title{Weak dipole moments of heavy fermions with flavor violation induced by $Z^\prime$ gauge bosons}

\author{J. Monta\~no-Dom\'inguez$^{1,2}$}
\author{B. Quezadas-Vivian$^{1}$}
\author{F. Ram\'irez-Zavaleta$^{1}$}
\author{E. S. Tututi$^{1}$}
\affiliation{
$^{1}$Facultad de Ciencias F\'isico Matem\'aticas, Universidad Michoacana de San Nicol\'as de Hidalgo,
Av. Francisco J. M\'ugica S/N, C.~P. 58060, Morelia, Michoac\'an, M\'exico.
\\
$^{2}$CONACYT, Av. Insurgentes Sur 1582, Col. Cr\'edito Constructor, Alc. Benito Ju\'arez,
C.~P. 03940, CDMX, M\'exico.
}

\begin{abstract}
A calculation of  weak dipole moments of charged fermions of the Standard Model (SM)  at the one-loop level, in the context of a general effective extended  neutral current model with flavor changing $Z^\prime f_i \bar{f}_j$ vertices, is presented. We give numerical predictions for the anomalous weak magnetic dipole moment (AWMDM) $a^w_{fi}$, and the weak electric dipole moment (WEDM) $d^w_{f_i}$, for the $\tau$ lepton and $t$ quark.
For  several  $Z^\prime$ gauge bosons considered, we find that, for the $\tau$ lepton, the best prediction for the real part of $a^w_{\tau}$ is of the order of $10^{-9}$, while the imaginary part is four orders of magnitude below. The highest value for the WEDM, $d^w_{\tau}$, corresponds to $10^{-26}$ $e$-cm, for its real part, and the imaginary part is  three orders of magnitude below. On the other hand, we found for the top quark, that the best prediction for the real part of $a^w_t$ is of the order of $10^{-7}$ and its imaginary part is of the order of  $10^{-11}$. We also found that $d^w_t$ is of the order of $10^{-26}$ $e$-cm for its real part, and its imaginary part can be as high as $10^{-31}$ $e$-cm.

\end{abstract}



\maketitle

\section{Introduction}
\label{sec:intro}

Magnetic dipole moment (MDM) and electric dipole moment (EDM)  are physical properties of elementary particles that have received much attention for a long time. In fact, since the seminal work by Schwinger on the magnetic dipole moment anomaly of the electron  \cite{schwinger}, physicists have dedicated much effort   in the  calculation  and the set up  of experiments of precision test  for  these  kind of electromagnetic properties of particles \cite{kinoshita2018}.  Due to the spin  in fermions,   electric dipole moment results  very important in the searching for violation of the  Conjugation and Party (CP) symmetry. Indeed, for instance,  a non-vanishing value for the EDM of the electron, implies a non conservation of the CP symmetry \cite{zeldovich1960,pais-primack,bernreuther,cairncrossetal}.
On the other hand, less attention has been paid on the weak electric dipole moment (WEDM) and the anomalous weak magnetic dipole moment (AWMDM)  in fermions, which arise in the Standard Model (SM) at the one-loop level. Nevertheless, weak dipole moments  can be of worth in the quest for new physics (NP) effects through  contributions resulting from quantum fluctuations to the couplings $Zf\bar{f}$ (where $f$ denotes   $\mu,\tau$ leptons or $t, b$ quarks and $Z$ represents the massive  neutral gauge boson), that are already  comprised in the SM at the tree level. For the case of leptons, the mentioned  couplings have been of interest in precision measurements. The latest experimental results  on the WEDM for the $\tau$ lepton,  were reported by the OPAL and ALEPH collaborations, the   bounds   are $|d^{w}_\tau| < 7.0 × 10^{-17}$ $e$-cm with a 95\% and $|d^{w}_\tau| < 3.7 × 10^{-17}$ $e$-cm with a 95\% confidence level,  respectively \cite{opal-92,aleph-92}.  Later, ALEPH Collaboration, by analyzing data collect from the reaction $e^+e^-\to \tau^+\tau^-$ at energies close to the $Z$ mass, measured both the real and imaginary components of the AWMDM and the CP-violating WEDM \cite{aleph-2003}. With  an integrated luminosity of 155 pb$^{-1}$, the  values reported are as follows: for the tau  lepton  the AWMDM is such that Re$(a_\tau^{w}) < 1.1 \times 10^{-3}$ at 95\% C.L., and Im$(a_\tau^{w}) <2.7\times 10^{-3}$ at 95 \% C.L., while,  the value for the  WEDM is given by  Re$(d_\tau^{w}) < 0.5\times10^{-17} e$-cm at 95\% C.L., and Im$(d_\tau^{w}) < 1.1 \times ^{-17} e$-cm at 95\% C.L. The experimental bounds are  far from the current  capabilities required to test the respective SM predictions, however these capabilities will  be increased enough to be at the reach of detection in the future. In relation to the $t$ quark and their  weak dipole moments, there are only theoretical predictions reported in the literature.  Within the SM context, the value reported of the contribution to the AWMDM of the top quark at large momentum transfer $\sqrt{q^2}=500$ GeV is $a_t^w(SM)=(-5_\cdot 60 + i~ 5_\cdot 35)\times 10^{-3}$ \cite{Bernabeu:1995gs}.
The prediction of the AWMDM for the $\tau$ lepton according to the SM is $a_\tau^w(M_Z^2) = - (2.10 + 0.61 i)\times 10^{-6}$ \cite{Bernabeu:1994wh}. Other works  in different models have also been obtained for  the weak dipoles moments of the muon and tau leptons and  heavy  quarks  \cite{Bernabeu:1995gs,Bernabeu:1997je,Hollik:1997vb,deCarlos:1997br,Hollik:1997ph,GomezDumm:1999tz,Bolanos:2013tda,Arroyo-Urena:2016ygo,Arroyo-Urena:2017sfb,Arroyo-Urena:2015uoa,Hollik:1998vz,Moyotl:2012zz}.  In particular, this work addressees the study of the AWMDM and WEDM within the context of flavor violation due to a massive neutral gauge boson.

Flavor violation is a topic of current interest, since it presents several  features that allow the search for  NP beyond the SM \cite{buras,feruglio}. In the quark sector of the SM,  flavor transitions produced by neutral currents  are induced at the loop level and, due to the GIM mechanism, these transitions are very suppressed \cite{SMqsectorsup}. For the lepton sector, the corresponding SM Lagrangian contains the exact flavor  symmetry, which implies that the leptonic number is conserved separately, that is to say, transitions between charged leptons  are forbidden. Nonetheless, neutrino oscillations \cite{Fukuda:1998mi} suggest that the lepton flavor symmetry is not conserved, since massive neutrinos  allow, for instance, the $\mu\to e\gamma$ decay in the minimal extended Standard Model \cite{cheng-li}. However the symmetry in question might be satisfied globally \cite{bilenky,feruglio}. Transitions between fermions of different flavor can be enhanced through flavor changing neutral currents (FCNC)  mediated  by a massive neutral gauge boson, identified in a generic manner as $Z^\prime$. As a matter of fact, if this gauge boson does exist, FCNC are allowed at the tree level. The existence of this boson has been proposed in numerous extended models, being  the simplest ones those  that involve an extra $U^\prime(1)$ gauge symmetry group \cite{langacker1}. Certainly, the simplest model  is based on the $SU_L(2)\times U_Y(1)\times U^\prime(1)$ extended electroweak gauge group \cite{robinett,langacker2,leike,perez-soriano}. There are experimental results on the search for the $Z^\prime$ gauge boson  predicted in different models.   The ATLAS Collaboration reports, with a $95\%$ confidence level in proton-proton collisions at $\sqrt{s}= 13$ TeV, minimal bounds ranging from 3.8 to 4.5 TeV, depending on the model~\cite{ATLAS2017}. The CMS Collaboration~\cite{CMS2018} found lower bounds that,  also depending on the model, going from 3.9 up to 4.5 TeV.

As already mentioned,  the AWMDM and  WEDM  of massive fermions have been studied in different contexts. Apart from the SM, the weak dipoles have also been studied in:  the Minimal Supersymmetric Standard Model (MSSM) \cite{Hollik:1997vb, deCarlos:1997br, Hollik:1997ph}, two-Higgs doublet models (THDMs) \cite{Bernabeu:1995gs, GomezDumm:1999tz, Arroyo-Urena:2015uoa}, leptoquarks \cite{Bolanos:2013tda}, the simplest little Higgs model \cite{Arroyo-Urena:2016ygo}, models with an extended scalar sector \cite{Arroyo-Urena:2017sfb}, supersymmetric theories \cite{Hollik:1998vz}, and unparticles \cite{Moyotl:2012zz}.
However, up to our knowledge, it does not exist in the literature, studies of the AWMDM and WEDM of massive fermions, in the context of FCNC induced by a $Z^\prime$ gauge boson.  In fact, the study of these weak dipoles in the mentioned context  comes to complement the   literature existent on this topic and it could open another window in the comprehension of transitions that violate the flavor symmetry. Moreover the latest experimental reports on the bounds of the mass of  $Z^\prime$ gauge bosons  renewals interest on the physics that could emerge at the TeV scale  involving these particles. As matter of fact, in this new era of precision measurements, possible deviation due to effects of new physics coming from physics beyond the SM   could be  perceptible and hence important.

In this article, we present the study of  weak dipole moments of the tau lepton and top quark within  the context of a general  flavor changing neutral currents  mediated by a $Z^\prime$ in the $SU_L(2)\times U_Y(1)\times U^\prime(1)$ model, with and without the CP symmetry. To do that, we use various   extended models that predict a $Z^\prime$ and adjust the respective coupling employed in the calculation of the transition amplitude and the weak dipole moments.

The outline of the paper is organized as follows. In Section II, the basic of FCNCs induced by a new neutral massive gauge boson of spin 1 is presented, and it is explained how bounds over $Z^\prime f_if_j$ (for $f_if_j= \tau u, \tau e,  tc,  tu$) couplings are determined.  In Section III, we exhibit the  analytical results for the electromagnetic weak dipole moments induced by FCNCs. In Sec. IV we present numerical predictions for the AWMDM and the WEDM of the tau lepton and top quark. Finally, we present our conclusions in Section V.

\section{Basic Formalism}

In order to carry out the calculation of  weak dipole moments of heavy fermions within the context of  FCNC, a general   structure of the leptonic $Zf\bar{f}$ coupling is required. Accordingly, the most general Lorentz structure of the vertex function that couples a $Z$ gauge boson with  fermions on shell, in terms of independent form factors,  can be expressed as follows  \cite{Hollik:1998vz,nowakowski}:
\begin{equation}
\Gamma_\mu^{Zff}(q^2)=\left(V^Z_f(q^2)+A^Z_f(q^2)\gamma_5\right)\gamma_\mu+ \left(F_{f\,M}^Z(q^2) -iF_{f\,E}^Z(q^2)\gamma_5\right)i\sigma_{\mu\nu}q^\nu
\label{general-vertex}
\end{equation}
where, as usual, $q=p-p^\prime$ is the  transfer momentum  and $\sigma^{\mu\nu}\equiv \frac{i}{2}[\gamma^\mu, \gamma^\nu]$. The  $V^Z_f(q^2)$ and $A^Z_f(q^2)$ form factors parameterize the  vector and axial currents; at the tree level,  they result constants and are  encompassed in the electroweak sector  of the SM.  The  $F_{f\,M}^Z(q^2)$ and $F_{f\,E}^Z$ form factors are related to the weak dipole moments of a charged fermion $f$ with mass $m_f$ as follows:
\begin{align}\label{Dipole-moments}
 a_f^w= -2 m_f F_{f\,M}^Z(m_Z^2), \qquad      d_f^w= -e F_{f\,E}^Z(m_Z^2),
\end{align}
expressions that will be used in what follows.

\subsection{The extended model}
Since it is required to estimate the strength of $Z^\prime f_if_j$  couplings, (where $f_{i}$ stands for any SM charged fermion) in order to determine their impact on the AWMDM and WEDM, it is necessary to establish the Lagrangian that comprises FCNC mediated by a $Z^\prime$ gauge boson. The most general renormalizable electroweak Lagrangian with the symmetry  $SU_L(2)\times U_Y(1)\times U^\prime(1)$ that includes FV mediated by a new neutral massive gauge boson, coming from extended  or grand unification models (GUT)\cite{robinett,langacker2,leike,perez-soriano,Robinett:1981yz}, is
\begin{align}
\mathcal{{L}}_{NC}&=\sum\limits_{i,j}\,\Bigg[\bar{f}_{i}\,\gamma^\alpha(\Omega_{Lfifj}\,P_L+\Omega_{Rfifj}\,P_R) f_j +
\bar{f}_{j}\,\gamma^\alpha (\Omega^{*}_{Lfifj}\,P_L+
\Omega^{*}_{Rfifj}\,P_R) f_i\Bigg] Z^{\prime}_{\alpha}, \label{LRG}
\end{align}
where the sum is extended over the fermions, $f_{i}$, of  the SM.  As usual, $P_{L}=\frac{1-\gamma_5}{2}$, ($P_{R}=\frac{1+\gamma_5}{2}$) represents  the chiral left (right) projector and $Z^\prime_{\alpha}$ is the neutral massive gauge boson predicted in several extensions of the SM.
The various $\Omega_{Lf_if_j}$ and $\Omega_{Rf_i f_j}$ parameters represent the strength of the $Z^\prime f_i f_j$ coupling. In what follows, we will assume that $\Omega_{Lf_i f_j}=\Omega_{Lf_j f_i}$ and $\Omega_{Rf_i f_j}=\Omega_{Rf_j f_i}$. The Lagrangian in Eq. \ref{LRG} includes both flavor-conserving and flavor-violating couplings mediated by a $Z^\prime$ gauge boson. In this work, the following $Z^\prime$ bosons are considered: the $Z^\prime_S$ of the sequential $Z^\prime$ model, the $Z^\prime_{LR}$ of the left-right symmetric model, the $Z^\prime_\chi$ boson that arises from the breaking of $SO(10) \rightarrow SU(5)\times U(1)$, the $Z^\prime_\psi$ that emerges as a result of $E_6 \rightarrow SO(10) \times U(1)$, and the $Z^\prime_\eta$ appearing in many superstring-inspired models \cite{langacker2,AYDE}. Concerning to the flavor-conserving couplings, $Q_{L,R}^{f_i}$ \cite{langacker2,Arhrib:2006sg,robinett}, the values of which are shown in Table \ref{cuadro}, for different extended models are related to the $\Omega$ couplings as $\Omega_{Lf_{i}f_{i}}= -g_2 Q^{f_i}_L$ and $\Omega_{Rf_{i}f_{i}}= -g_2 Q^{f_i}_R$, where $g_2$ is the gauge coupling of the $Z^\prime$ boson. For the extended models we are interested in, the gauge couplings of $Z^\prime$'s are
\begin{equation}
g_2=\sqrt{\frac{5}{3}} \sin \theta_W g_1 \lambda_g,
\end{equation}
where $g_1= g/ \cos \theta_W$, $\lambda_g$ depends on the symmetry-breaking pattern being of $\mathcal{O}(1)$ \cite{Robinett:1981yz}, and $g$ is the weak coupling constant. In the sequential $Z$ model, the gauge coupling $g_2=g_1$.
\begin{table}[!t]
\centering
\begin{tabular}{c c c c c c}
\hline\hline
            & $Z^\prime_{S}$ & $Z^\prime_{LR}$ &        $Z^\prime_{\chi}$      &      $Z^\prime_{\psi}$     & $Z^\prime_{\eta}$\\
\hline
$Q_{L}^{l_i}$ &-0.2684  & 0.2548   &$\frac{3}{2\sqrt{10}}$ &$\frac{1}{\sqrt{24}}$ &$\frac{1}{2\sqrt{15}}$ \\
$Q_{R}^{l_i}$ & 0.2316  &-0.3339   &$\frac{-3}{2\sqrt{10}}$&$\frac{-1}{\sqrt{24}}$&$\frac{-1}{2\sqrt{15}}$\\
$Q_{L}^{u_i}$ & 0.3456  & -0.08493 &$\frac{-1}{2\sqrt{10}}$&$\frac{1}{\sqrt{24}}$ &$\frac{-2}{2\sqrt{15}}$\\
$Q_{R}^{u_i}$ &-0.1544  & 0.5038   &$\frac{1}{2\sqrt{10}}$ &$\frac{-1}{\sqrt{24}}$&$\frac{2}{2\sqrt{15}}$ \\
\hline\hline
\end{tabular}
\caption{\small Chiral-diagonal couplings of the extended models.}\label{cuadro}
\end{table}

\section{One-loop $Z^\prime$ contribution to the WDM}
\begin{figure}[h!]
\begin{center}
\includegraphics[width=6 cm]{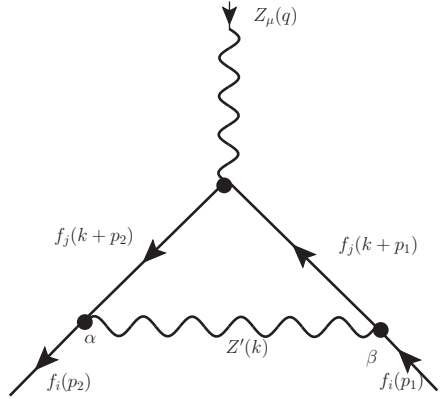}
\caption{\small {One-loop contribution to the static weak electromagnetic dipole moments induced by a $Z^\prime$ gauge boson with flavor violation.}} \label{dia1}
\end{center}
\end{figure}

Analytical results at the one-loop level  for the AWMDM and WEDM for any charged fermion of the SM, induced by FCNCs mediated by the $Z^\prime$ gauge boson are presented in this section. However, due to their possible interest in the search for NP beyond the SM,
we only carry out explicit numerical estimations of  weak dipole moments (WDMs) corresponding to the tau lepton and top quark.
For our purpose, it is convenient  to express the $ Z^\prime{f_{i}} f_{j}$ couplings in such a way that we can identify the vector ($g_{VZ^\prime}^{f_{i} f_{j}}$) and axial ($g_{AZ^\prime}^{f_{i} f_{j}}$) parameters of the couplings in a similar manner as for the $Zf\bar{f}$ coupling of the SM. To this end, we use the chiral projectors above defined and set
\begin{equation}
\Omega_{L f_{i}f_{j}}P_L + \Omega_{R f_{i}f_{j}}P_R = g_{VZ^\prime}^{f_{i} f_{j}} - g_{AZ^\prime}^{f_{i} f_{j}}\gamma^{5},
\end{equation}
to obtain
\begin{equation}
g_{VZ^\prime}^{f_{i} f_{j}} \equiv \frac{1}{2}(\Omega_{L f_{i}f_{j}} + \Omega_{R f_{i}f_{j}}), \hspace{1 cm}\hspace{1 cm} g_{AZ^\prime}^{f_{i} f_{j}} \equiv \frac{1}{2}(\Omega_{L f_{i}f_{j}} - \Omega_{R f_{i}f_{j}}).
\end{equation}
Along with the corresponding  values for the $g_{VZ}^{f_j}$ and $g_{AZ}^{f_j}$ parameters in the SM, which are listed in the Table \ref{valores SM}, the  $g_{VZ^\prime}^{f_{i} f_{j}}$ and $g_{AZ^\prime}^{f_{i} f_{j}}$ parameters will be used as inputs in our calculation.
{\Large \begin{table}[!t]
		\centering
		{\normalsize \begin{tabular}{c  c  c  c}
				\hline\hline
				$f_j$ & $Q_{f_j}$ &$g_{VA}^{f_j}$ & $g_{VZ}^{f_j}$\\
				\hline
				$\nu_{e}$, $\nu_{\mu}$, $\nu_{\tau}$  & $0$  &$\frac{1}{2}$ &$\frac{1}{2}$  \\
				
				$e$,  $\mu$,  $\tau$  &-1&-$\frac{1}{2}$& -$\frac{1}{2}$ + $2 \sin^2 \theta_W$\\
				
				$u$,  $c$,  $t$  & $\frac{2}{3}$ &$\frac{1}{2}$&$\frac{1}{2} - \frac{4}{3} \sin^2 \theta_W$ \\
				
				$d$,  $s$,  $b$ &-$\frac{1}{3}$&-$\frac{1}{2}$ &-$\frac{1}{2}$+$\frac{2}{3}\sin^2\theta_W$ \\
				\hline\hline
		\end{tabular}}
		\caption{\small Values of the $g_{VA}^{f_j}$ and $g_{VZ}^{f_j}$ parameters in the SM. }\label{valores SM}
\end{table}  }
The contribution of the $ Zf_i f_i$ couplings to the  weak electromagnetic dipole moments at the one-loop level induced by the $Z^\prime$ gauge boson, can be obtained from the invariant amplitude calculated in the unitary gauge (see Fig. \ref{dia1})
\begin{eqnarray}
\mathcal{M}_{f_i}
= -i^6\frac{g}{2 c_W}\sum_{f_j}\int\frac{d^4k}{(2\pi)^{4}}\bar{u}(p_2) [\gamma^{\alpha}(g_{VZ^{\prime}}^{f_{i}    f_{j}}-g_{AZ^{\prime}}^{f_{i}f_{j}}\gamma^{5})](\cancel{k}+ \cancel{p}_2+m_{f_{j}})\gamma^\mu (g_{VZ}^{f_j}-g_{AZ}^{f_{j}}\gamma^{5})
\nonumber \\
\times \frac{(\cancel{k}+\cancel{p_1}+m_{f_{j}})[\gamma^{\beta}(g_{VZ^{\prime}}^{f_{i}f_{j}*}-g_{AZ^{\prime}}^{f_{i}f_{j}*}\gamma^{5})]}{(k^2-m_{Z^{\prime}}^{2})[(k+p_2)^2-m_{f_{j}}^2][(k+p_1)^2-m_{f_{j}}^{2}]}
u(p_1)\left(-g_{\alpha \beta} + \frac{k_{\alpha }k_{\beta}}{m_{Z^{\prime}}^{2}}\right)\epsilon_\mu(q). \label{amplitud}
\end{eqnarray}
where the sum is taken over leptons or quarks in the loop, depending on the fermion  $f_i$.
The loop integral indicated in Eq. (\ref{amplitud}) is solved through  Passarino-Veltman tensor decomposition method using \texttt{FeynCalc} \cite{feyncalc}, and the resulting scalar functions were integrated by using the Feynman parametrization method and cross-checked with \texttt{Package-X} \cite{packageX}. After  lengthy algebraic manipulations and using Eqs. (\ref{general-vertex}) and (\ref{Dipole-moments}), we can express the AWMDM for a fermion $f_i$ as
\begin{equation}
\begin{split}
a_{f_i}^{w}  = \sum_{f_j}&\left[ g_{VZ}^{f_j} \left\{ |g_{VZ'}^{f_i f_j}|^2 F_V^a(m_{f_i}, m_{f_j}, m_Z, m_{Z'}) + |g_{AZ'}^{f_i, f_j}|^2F_A^a(m_{f_i}, m_{f_j}, m_Z, m_{Z'}) \right\}\right.
\\
&\left.  + g_{AZ}^{f_j} \left\{ g_{AZ'}^{f_i f_j}g_{VZ'}^{f_i f_j*}+g_{AZ'}^{f_i f_j*}g_{VZ'}^{f_i f_j} \right \}F_{VA}^a(m_{f_i}, m_{f_j}, m_Z, m_{Z'})\right], \label{WMDM}
\end{split}
\end{equation}
where
\[
|g_{VZ'}^{f_i f_j}|^2=
\frac{1}{4} \left[ (\text{Re}\Omega_{Lf_if_j}+\text{Re}\Omega_{Rf_if_j})^2
+(\text{Im}\Omega_{Lf_if_j}+\text{Im}\Omega_{Rf_if_j})^2 \right] \ ,
\]
\[
|g_{AZ'}^{f_i f_j}|^2=
\frac{1}{4} \left[(\text{Re}\Omega_{Lf_if_j}-\text{Re}\Omega_{Rf_if_j})^2
+(\text{Im}\Omega_{Lf_if_j}-\text{Im}\Omega_{Rf_if_j})^2 \right. ,
\]
and the $F_V^a$, $F_A^a$ and $F_{VA}^a$ form factors are explicitly displayed in the appendix. For the WEDM we obtain
\begin{equation}
\begin{split}
d_{f_i}^{w}= \sum_{f_j}g_{VZ}^{f_j} \left\{  g_{VZ'}^{f_i f_j}g_{AZ'}^{f_i f_j*}- g_{AZ'}^{f_i f_j}g_{VZ'}^{f_i f_j*}  \right\} F_{VA}^d(m_{f_i}, m_{f_j}, m_Z, m_{Z'}), \label{WEDM}
\end{split}
\end{equation}
where
\begin{equation}
g_{VZ'}^{f_i f_j}g_{AZ'}^{f_i f_j*}- g_{AZ'}^{f_i f_j}g_{VZ'}^{f_i f_j*} =
i \left( \text{Re}\Omega_{Lf_if_j}\text{Im}\Omega_{Rf_if_j}
-\text{Re}\Omega_{Rf_if_j}\text{Im}\Omega_{Lf_if_j} \right)\ ,
\label{condition}
\end{equation}
and the $F_{VA}^d$ form factor is  also presented in the appendix. Let us mention that we have verified that the contributions to both the AWMDM and WEDM from the diagram of Fig.\ref{dia1} are free of ultraviolet divergences. In addition,  if we take the $Z^\prime=Z$ gauge boson along with the couplings replaced by the corresponding of the SM, we can corroborate that the expression in Eq. (\ref{WMDM}) reproduces the results for the contributions of the $Z$ gauge boson to the SM, computed in \cite{Bernabeu:1995gs} for the $Z\tau\tau$ contribution, whose value is $a_{\tau}^w=4.13\times10^{-8} + 1.91 i \times10^{-8}$.  Let us mention that  similar expressions to Eqs. (\ref{WMDM}) and (\ref{WEDM})  were presented in Ref. \cite{Hollik:1997ph} within the context of the minimal super-symmetric model, which are in agreement with ours. Also in Ref. \cite{muon-g-2}  can be found  similar results, but for the $g-2$ value of the muon lepton.

Even though Eqs. (\ref{WMDM}) and (\ref{WEDM}) for the weak magnetic and electric dipolar moments respectively are quite general, we  restrict the discussion only for the cases of the top quark and  tau lepton. This relies on the fact that there are only experimental results concerning the weak dipoles  for the tau  lepton, such that we can compare our numerical results with the  experimental ones. Although, as already mentioned, there are several studies of the weak dipole moments for the top quark and tau lepton in different theoretical contexts, no studies have been presented on the context of FCNC mediated by a massive neutral gauge boson. That also allows us to compare the respective numerical results. On the other hand, a Taylor expansion of  the form factors involved in the weak dipoles moments around  the $m_{fi}=0$ shows that dominant terms are  $F_{V,A}^{a,d}\propto\left(\frac{m_{fi}}{M_{Z'}}\right)^2$. Hence,  the various $F_{V,A}^{a,d}$ for lighter fermions  are even more suppressed. Moreover, for the case of light quarks, their masses and couplings involved must be substituted by the running masses and couplings in order to obtain more accurate results.

For convenience and future discussion, we split explicitly the  different contributions to the AWMDM and WEDM of the fermion in question, which can be read off from Eqs. (\ref{WMDM}) and (\ref{WEDM}), respectively. For the tau lepton, we obtain
\begin{eqnarray}
a_\tau^w &=& a_{\tau e}^w + a_{\tau \mu}^w + a_{\tau \tau}^w,\nonumber\\
d_\tau^w &=& d_{\tau e}^w + d_{\tau \mu}^w + d_{\tau \tau}^w,
\label{contrubitionatau}
\end{eqnarray}
where, for example
\begin{eqnarray}
a_{\tau e}^w&=&g_{VZ}^{e} \left( |g_{VZ'}^{ \tau e }|^2 F_V^a(m_{\tau},m_e, m_Z, m_{Z'}) + |g_{AZ'}^{\tau e}|^2F_A^a(m_{\tau}, m_{e}, m_Z, m_{Z'}) \right)\nonumber\\
&& + g_{AZ}^{e} \left\{ g_{AZ'}^{ \tau e }g_{VZ'}^{\tau e*}+g_{AZ'}^{\tau e *}g_{VZ'}^{\tau e} \right \}F_{VA}^a(m_{\tau}, m_{e},  m_Z, m_{Z'}).
\end{eqnarray}
For the top quark, we have
\begin{eqnarray}
a_t^w &=& a_{t u}^w + a_{t c}^w + a_{t t}^w,\nonumber\\
d_t^w &=& d_{t u}^w + d_{t c}^w + d_{t t}^w.
\label{contrubitiontop}
\end{eqnarray}
With similar expressions  for the different contributions, as the given  for the tau lepton.
Let us analyze the weak dipole moments, according to their  CP-symmetry properties, depending whether $d_t^w$ vanishes or not. The form factors are  non vanishing as it can be observed in the appendix A. Thus, we can identify  the following   scenarios
\begin{enumerate}
\item The CP conserving (CP-C) case. For this instance, $d_{f_i}^w$=0, which is satisfied when either $\Omega_{Lf_if_j}=0$, or $\Omega_{Rf_if_j}=0$, or $Re \Omega_{Lf_if_j}=Re \Omega_{Rf_if_j}=0$, or $Im \Omega_{Lf_if_j}=Im \Omega_{Rf_if_j}=0$, which it occurs for  $f_i=\tau$ ($f_j=e,\mu$ and  $\tau$) or $f_i=t$ ($f_j=u,c$ and $t$).  For each case, the  other  elements are maintained different  from zero. In general, the CP-C case emerges whenever $g_{VZ^\prime}^{f_i f_j}g_{AZ^\prime}^{f_i f_j*}- g_{AZ^\prime}^{f_i f_j}g_{VZ^\prime}^{f_i f_j*} =0$.
In order to compute the AWMDM for the case of CP conserving, we have numerically tested  different combinations for the couplings $\Omega_{R,L}$  concluding that the results are, in essence, the same. Thus, we will consider, for the calculation presented below, the case in which $\Omega_{Lf_if_j}\neq 0$ and $\Omega_{Rf_if_j}= 0$.

\item The CP violating (CP-V) case.  Both $a_{f_i}^w$ and $d_{f_i}^w$, can occur. In this instance,  we restrict ourselves to the case of, $\text{Re}\Omega_{Lf_if_j}\neq 0$ and $\text{Im}\Omega_{Rf_if_j}\neq 0$, keeping the remaining  terms in Eq. (\ref{condition}) equal to zero.
	
\end{enumerate}

\section{Numerical analysis}
\subsection{AWMDM  of the tau lepton}
Here, we perform the numerical analysis on the weak electromagnetic moments of the  tau lepton. To this end, we resort to  Eq. (\ref{WMDM}) for the tau AWMDM, and the Eq. (\ref{WEDM}) for the tau WEDM, by considering different $Z^\prime$ gauge bosons models, namely, $Z^\prime_{S}$, $Z^\prime_{LR}$ ,$Z^\prime_{\chi}$, $Z^\prime_{\psi}$ and $Z^\prime_{\eta}$. The weak dipoles rely on the different couplings derived according to the models above mentioned. The diagonal elements of the $\Omega_{R, L}$ matrices to be used, can be obtained in terms of the chiral charges shown in  Table \ref{cuadro}. The off-diagonal elements, such as $\Omega_{L,R \tau e}$ and $\Omega_{L,R \tau \mu}$,  were estimated in Ref. \cite{Aranda:2012qs,Aranda:2018zis} and they will be used in the numerical calculation of the WEDM and AWMDM. The other inputs are the masses of the fermions  as well as the couplings  contained in Table \ref{valores SM} whose numerical values were taken from the PDG \cite{pdg-zyla-etal}.

\subsubsection{CP conserving case}
\begin{center}
	\begin{figure}[t!]
		\includegraphics[scale=0.45]{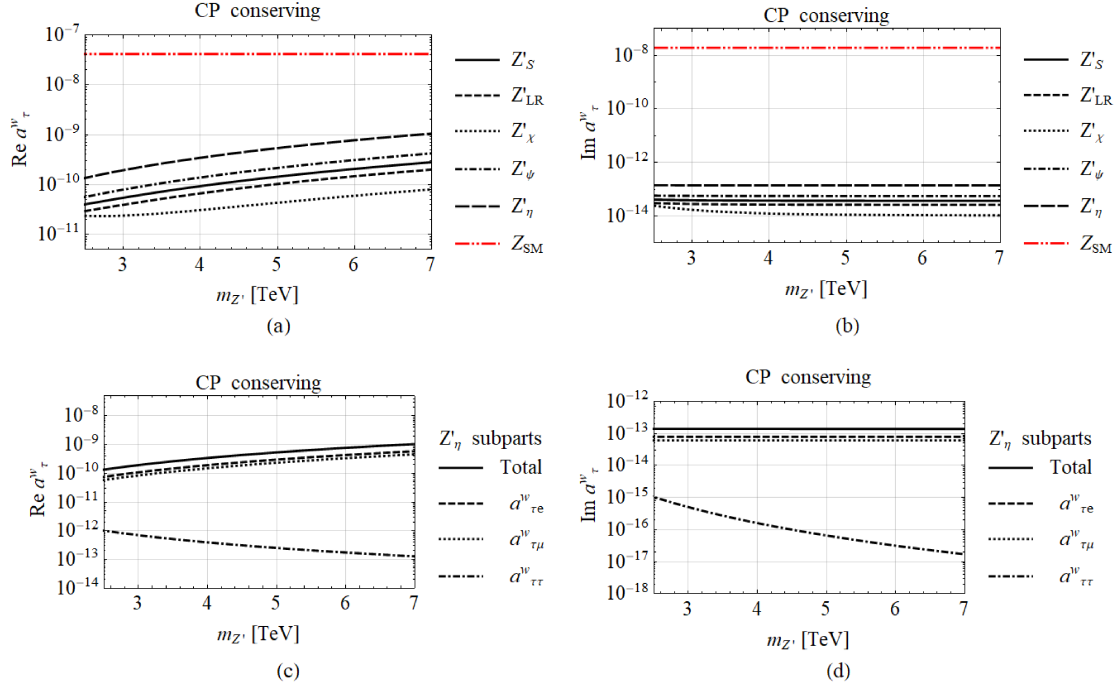}
		\caption{\small {Anomalous weak magnetic dipole moment of the tau lepton with CP conservation, induced by a $Z^\prime$ gauge boson with flavor violation. (a) Contributions of the $Z^\prime$  bosons from different models to Re ${a^w}_{\tau}$ and (b) the Im ${a^w}_{\tau}$. (c), (d) The respective real and imaginary parts generated by the subparts of the main contribution due to $Z^\prime_{\eta}$ boson.}}\label{tauccpm}.
	\end{figure}
\end{center}

Let us  first analyze the  CP-C case.  In  Fig. \ref{tauccpm}, it is shown the $a_\tau^w$ as a function of the $Z^\prime$ gauge boson mass in the interval $m_{Z^\prime}= [2.5, 7]$ TeV.
The Fig. \ref{tauccpm}(a) shows  $\mathrm{Re}\,a_\tau^w$, where it can be appreciated  the contribution from  various $Z^\prime$ gauge bosons, the intensities of these contributions go from $10^{-11}$ to $10^{-9}$. We can also observe that, as the mass of the boson $Z^\prime$ is increased, the intensity of $\mathrm{Re}\,a_\tau^w$ grows rather slow, along the mass interval. The highest value is provided by the $Z^\prime_\eta$ boson.  Notice that for values of $m_{Z^\prime}>5$ TeV is barely one order of magnitude below from the SM $Z$ gauge boson contribution, whose value is $\mathrm{Re}\,a_\tau^w(m_Z) = 4.13142 \times 10^{-8}$, shown in red line.  This value is well below from the current  experimental bound Re$(a_\tau^w) < 1.1 \times 10^{-3}$ at 95\% C.L. \cite{aleph-2003}. The lowest signal is provided by $Z^\prime_\chi$ gauge boson, being of order $10^{-11}$.
The imaginary part of the AWMDM  ($\mathrm{Im}\,a_\tau^w$) is illustrated in Fig. \ref{tauccpm}(b). The contributions of the different $Z^\prime$ gauge bosons range between $10^{-13}$ and $10^{-14}$. Again,  the highest signal is provided by the $Z^\prime_\eta$ boson, and is five orders of magnitude below the $Z$ gauge boson contribution to the SM, shown in red line, whose numerical value is $\mathrm{Im}\,a_\tau^w(m_Z) = 1.9121\times 10^{-8}$. The current  experimental bound is $\mathrm{Im}\,a_\tau^w <2.7\times 10^{-3}$ at 95 \% C.L. \cite{aleph-2003}. The lowest signal is given by $Z^\prime_\chi$ boson being of the order of $10^{-14}$.
Let us highlight the case of the $Z^\prime_\eta$ gauge boson.  In Fig. \ref{tauccpm}(c), the real subparts of the main contribution belonging to this boson are illustrated in detail. The $a_{\tau e}^w$ and $a_{\tau \mu}^w$ parts are of the same order of magnitude, between $10^{-10}$ and $10^{-11}$, while $a_{\tau \tau}^w$ is two orders of magnitude below. In  Fig. \ref{tauccpm}(d) we can appreciate the imaginary subparts to the main contribution. Notice that values of the $a_{\tau e}^w$ and $a_{\tau \mu}^w$ parts are of the order of $10^{-12}$ and $10^{-13}$, respectively. In contrast, the $a_{\tau \tau}^w$  contribution is between 2 and 4 order of magnitude smaller.

In order to contextualize our results, let us compare them with the respective predictions of $a_\tau^w$ in  extended models. The estimations for $a_\tau^w$ coming from the simplest little Higgs model are of the order of $10^{-9}$ for the real part, and $10^{-10}$ for its imaginary part \cite{Arroyo-Urena:2016ygo}. As far as $a_\tau^w$ is concerned, in models with an extended scalar sector, its real part reaches values as high as $10^{-10} - 10^{-9}$, whereas the imaginary part is one or two orders of magnitude below \cite{Arroyo-Urena:2017sfb}. The predictions for the Minimal Supersymmetric Standard Model (MSSM) are for $\mathrm{Re}\,a_\tau^w$ of $10^{-6}$ and $\mathrm{Im}\,a_\tau^w$ of $10^{-7}$ \cite{deCarlos:1997br,Hollik:1997vb}. For the model of Unparticle Physics, the predictions is of $10^{-9}$ for the real and imaginary part of $a_\tau^w$ \cite{Moyotl:2012zz}, and in the THDMs for $\mathrm{Re}\,a_\tau^w$ is of the order $10^{-10}$ \cite{Bernabeu:1994wh}.

\subsubsection{CP violating case}
\begin{center}
	\begin{figure}[t!]
		\includegraphics[scale=0.5]{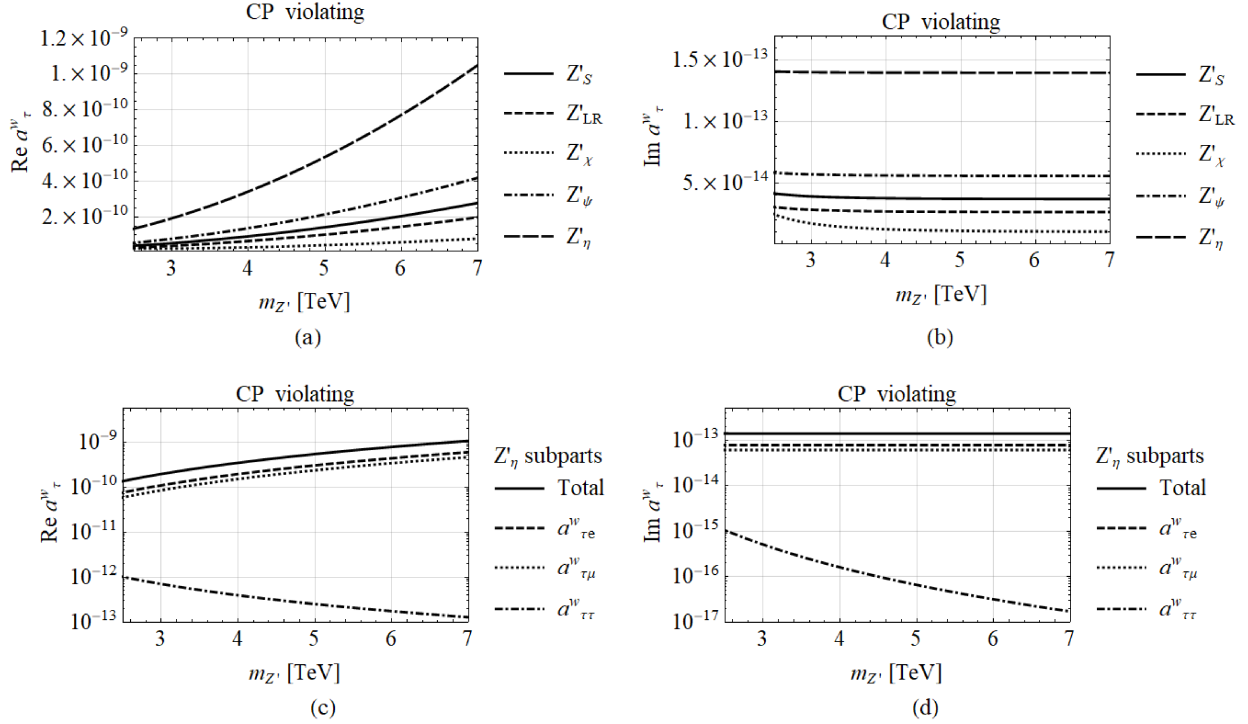}
		\caption{\small {Anomalous weak magnetic dipole moment of the tau lepton with CP violation, induced by a $Z^\prime$ gauge boson with flavor violation. (a) Contributions of the $Z^\prime$ gauge bosons from different models to Re ${a^w}_{\tau}$ and (b) the Im ${a^w}_{\tau}$. (c), (d) The respective real and imaginary parts generated by the main contribution due the $Z^\prime_{\eta}$ and its subparts.}}\label{tauvcpm}
	\end{figure}
\end{center}

In  Fig. \ref{tauvcpm}, it can be observed the behavior of the real and imaginary parts of the AWMDM for the CP violating case as a function of the mass of the various $Z^\prime$ gauge bosons in the interval $m_{Z^\prime} =[2.5,7]$ TeV. Once again,  the $Z^\prime_\eta$ gauge boson offers the  highest contribution to $\mathrm{Re}\,a_\tau^w$,  the values go from $10^{-10}$ to $10^{-9}$. The smallest contribution to  the weak anomaly is provided by the $Z^\prime_\chi$ gauge boson. As far as the behavior of the imaginary part of $a_\tau^w$ is concerned,  we can appreciate in Fig. \ref{tauvcpm} (b)  that $Z^\prime_{\eta}$ provides the highest signal, and that it is 4 orders of magnitude smaller than the real part.  The smallest value   corresponds to the $Z^\prime_{\chi}$ gauge boson, being  of the order of   $10^{-14}$.
In Fig. \ref{tauvcpm} (c), it is shown the subpart contributions belonging to the $Z^\prime_\eta$ gauge boson, as well as the total contribution due to  $a_{\tau e}^w$,  $a_{\tau \mu}^w$ and  $a_{\tau \tau}^w$ parts. We can observe that the main contributions are due to $a_{\tau e}^w$,  $a_{\tau \mu}^w$. In Fig \ref{tauvcpm} (d) it is displayed the imaginary part of   $a_\tau^w$ aside from the  contributions of different  subparts. The resulting value of the $\mathrm{Im}\,a_\tau^w$ is  of the order of $10^{-13}$, for the dominant contributions.

\subsection{The  WEDM of the tau lepton}
\begin{center}
	\begin{figure}[t!]
		\includegraphics[scale=0.5]{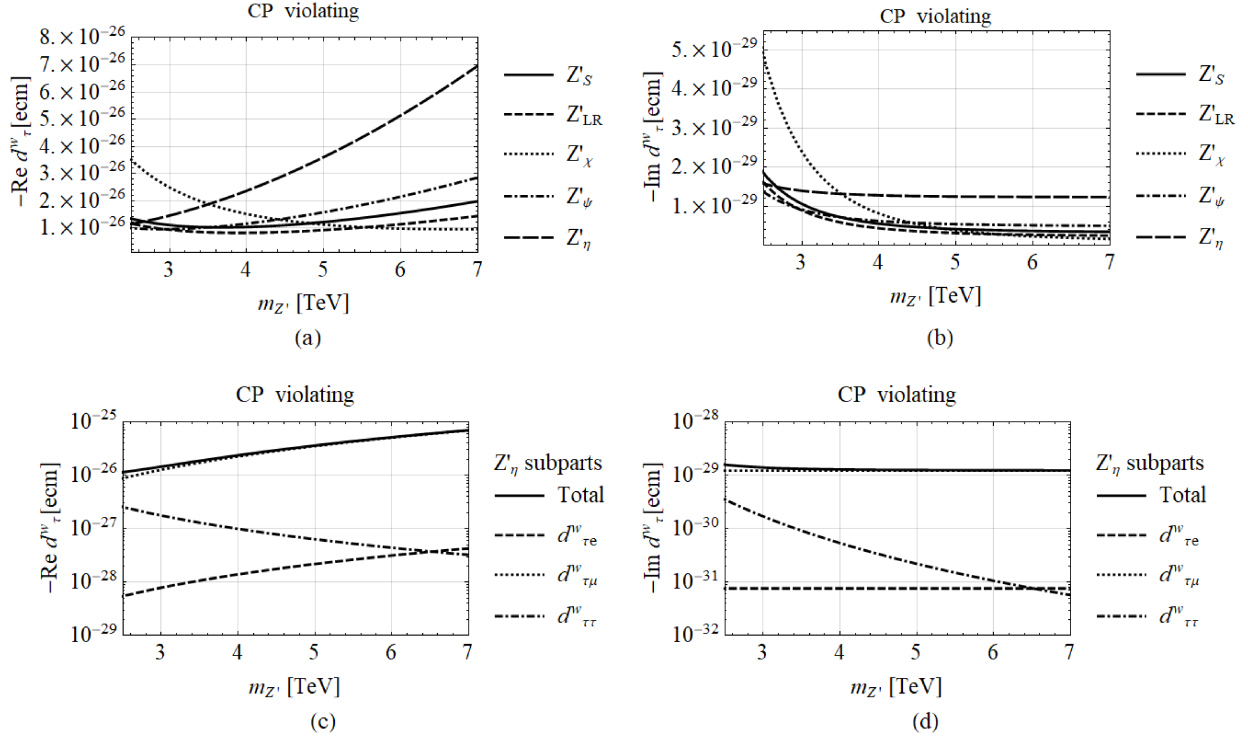}
		\caption{\small {Weak electric dipole moment of the tau lepton with CP violation, induced by a $Z^\prime$ gauge boson with flavor violation. (a) Contributions of the $Z^\prime$ gauge bosons from different models to Re ${d^w}_{\tau}$ and (b) the Im ${d^w}_{\tau}$. (c), (d) The respective real and imaginary parts generated by the main contribution due the $Z^\prime_{\eta}$ and its subparts.}}\label{tauvcpe}
	\end{figure}
\end{center}

In this subsection, we discuss the numerical results for  WEDM of the $\tau $ lepton. As usual, we express the value of  $d_\tau^w$ in $e$-cm units. In  Fig. \ref{tauvcpe}, it is displayed the different contributions to $d_\tau^w$ as a function of the $Z^\prime$ gauge boson mass,  in the interval $m_{Z^\prime} =[2.5,7]$ TeV. The Fig. \ref{tauvcpe} (a) shows the contributions of  the real part to $-d_\tau^w$ coming from  the $Z^\prime$ gauge bosons in question. As it can be observed, the values are of the order of $10^{-26}e$-cm. Again, the strongest prediction corresponds to the $Z^\prime_\eta$ gauge boson. The lowest value is offered by $Z^\prime_{LR}$. As already commented, the experimental bound for the real part of  WEDM of the $\tau$ lepton is  $\mathrm{Re}\,d_\tau^w < 0.5\times10^{-17} e$-cm at 95\% C.L. \cite{aleph-2003}.
The behavior of the  imaginary part of $-\mathrm{Im}\,d_\tau^w$ is illustrated in Fig. \ref{tauvcpe}(b). It can be  observed that  the  contributions due to the  $Z^\prime$ gauge boson  are of the same order, being around of $10^{-29}e$-cm. These  values are much smaller than the  current experimental bound for the imaginary part of  WEDM of the $\tau$ lepton, which is  $\mathrm{Im}\,d_\tau^w < 1.1 \times 10 ^{-17} e$-cm at 95\% C.L. \cite{aleph-2003}.
Finally, in  Figs. \ref{tauvcpe} (c) and (d), are shown the $\mathrm{Re}\,d_\tau^w$ and  $\mathrm{Im}\,d_\tau^w$ contributions of the  subparts  to the main prediction owing to the $Z^\prime_\eta$ gauge boson, respectively.  In both cases, the subpart $d_{\tau\mu} ^w$ gives the highest contribution to the imaginary part of the  WEDM of the $\tau$ lepton.

Let us now compare our results  with other theoretical predictions. In multi–Higgs models context \cite{Bernreuther:1992dz}, the corresponding value predicted  is  Re($d_\tau^w) \sim 3\times 10^{-22}e$-cm. In the case of leptoquark models \cite{Bolanos:2013tda}, the value found is $d_\tau^w \sim 10^{-19}e$-cm. On the other hand, models with an extended scalar sector predict   that the real part of $d_\tau^w$ is of the order of $10^{-24}e$-cm and its imaginary part can reach the intensity of  $10^{-26}e$-cm  \cite{Arroyo-Urena:2017sfb}. The  Minimal Supersymmetric Standard Model (MSSM) gives for $d_\tau^w$ a value of  order of  $10^{-21}e$-cm \cite{Hollik:1997ph}. For the model of Unparticle Physics the value predicted   is of $10^{-24}e$-cm for the real and imaginary part of $d_\tau^w$ \cite{Moyotl:2012zz}.   In the three doublet Higgs models, the value estimated for  $\mathrm{Re}\,d_\tau^w$ is of the order $10^{-22}e$-cm \cite{Bernabeu:1995gs}. These results, for the WEDM of the $\tau$ lepton, are much more larger  than our estimates or similar in some cases.

\subsection{AWMDM  of the top quark}

Now, we carry out the  analysis of the  weak electromagnetic dipole  moments for the top quark. To this end, we resort to  Eqs. (\ref{WMDM}) and (\ref{WEDM}) for the AWMDM, and  WEDM of the  quark in question, respectively.
As we can see from  Eq. (\ref{contrubitiontop}), the weak moments receive contributions from the coupling parameters $\Omega_{L,R t u}$, $\Omega_{L,R t c}$ and $\Omega_{L,R t t}$, that were previously  computed in Ref. \cite{Aranda:2020wfd}; they are considered as inputs for the calculation.
\subsubsection{AWMDM with CP conservation}

\begin{center}
	\begin{figure}[t!]
		\includegraphics[scale=0.5]{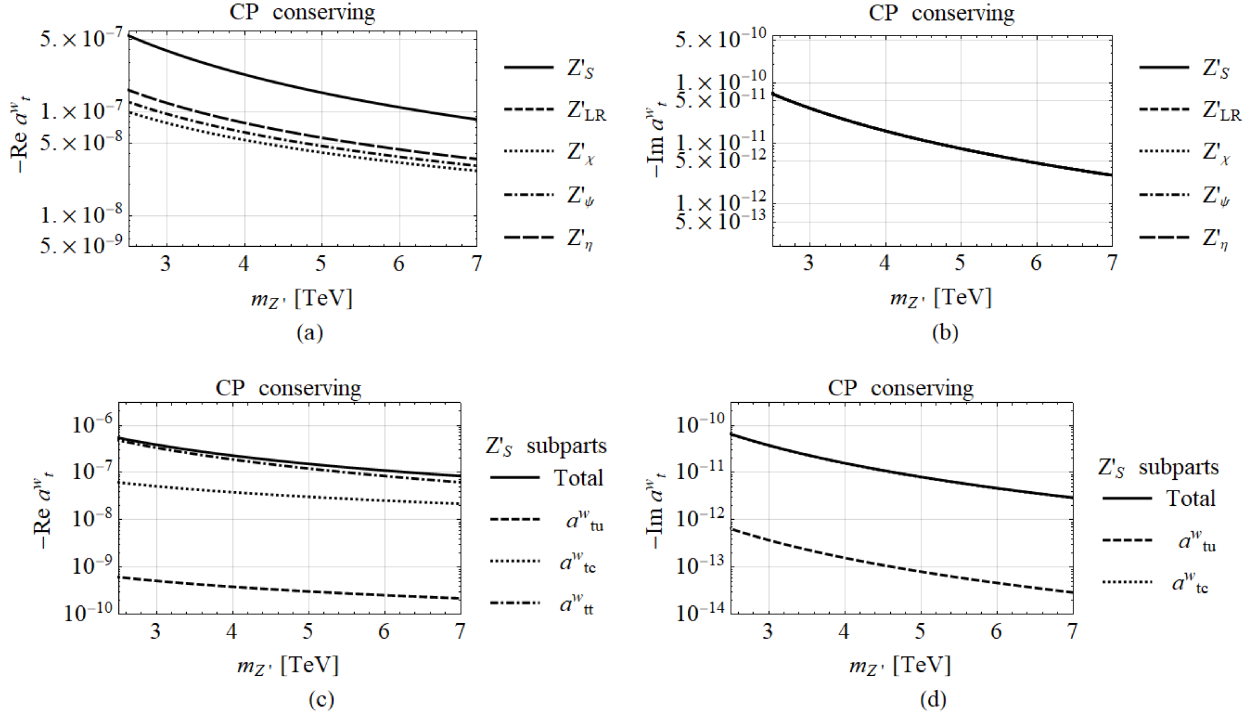}
		\caption{\small {Anomalous weak magnetic dipole moment of the top quark with CP conservation, induced by a $Z^\prime$ gauge boson with flavor violation. (a) Contributions of the $Z^\prime$ gauge bosons from different models to Re ${a^w}_{t}$ and (b) the Im ${a^w}_{t}$. (c), (d) The respective real and imaginary parts generated by the subparts of the main contribution due to $Z^\prime_{S}.$}}\label{topccpm}
	\end{figure}
\end{center}

In  Fig. \ref{topccpm} it is shown the numerical results for the AWMDM of the top quark as a function  of the $Z^\prime$ gauge boson mass, in the interval $m_{Z^\prime}= [2.5, 7]$. Let us remark that the values found for $a_t^w$ are in agreement with the current experimental bounds \cite{pdg-zyla-etal}. In Fig. \ref{topccpm} (a) the contributions due to different $Z^\prime$ gauge bosons to  $\mathrm{Re}\,a_t^w$  are presented. The numerical values go from $10^{-8}$ to $10^{-7}$ along  the interval used. The main contribution to $\mathrm{Re}\,a_t^w$ comes from the $Z^\prime_S$ gauge boson;  the $Z^\prime_\chi$ boson contributes with the smallest values.
The imaginary part of $a_t^w$ is illustrated in Fig. \ref{topccpm} (b). It is notorious that
all the different $Z^\prime$ bosons share essentially the same imaginary value, and they are three orders of magnitude smaller than the real part. In  Fig. \ref{topccpm} (c), the real subparts of the main contribution coming from the $Z^\prime_S$ gauge boson are depicted, with $a_{tt}^w$ being the highest one, while $a_{tu}^w$ presents the least value.
In Fig. \ref{topccpm} (d) we can appreciate the imaginary subparts of the main contribution due to $Z^\prime_S$ boson, which are generated  by the off-diagonal subparts  $a_{tc}^w$, which is of order  $10^{-11}$, and $a_{tu}^w$,  being of order  $10^{-13}$. The $a_{tt}^w$ contribution is negligible.

The value of  $Z$ gauge boson contribution to the top quark  AWMDM  in the SM,  at  the   momentum transfer  $\sqrt{q^2}=500$ GeV,   is $a_t^w=-2_\cdot 46\times 10^{-4}  -i~ 1_\cdot 45\times 10^{-3}$ \cite{Bernabeu:1995gs}. On the other hand, our calculations, showed in Figs. \ref{topccpm}, \ref{topvcpm} and \ref{topvcpe}, were performed at  $q^2=m_Z^2$.  However, in order to compare our prediction with the value found in  Ref. \cite{Bernabeu:1995gs}, we also did the calculations at the momentum transfer $\sqrt{q^2}=500$ GeV, finding that the values for the real parts are three orders of magnitude below from the $Z$ gauge boson contribution to the SM value, while our imaginary parts  are two or up to three orders of magnitude smaller, respectively.

\subsubsection{AWMDM of the top quark with CP violation}

\begin{center}
	\begin{figure}[h!]
		\includegraphics[scale=0.5]{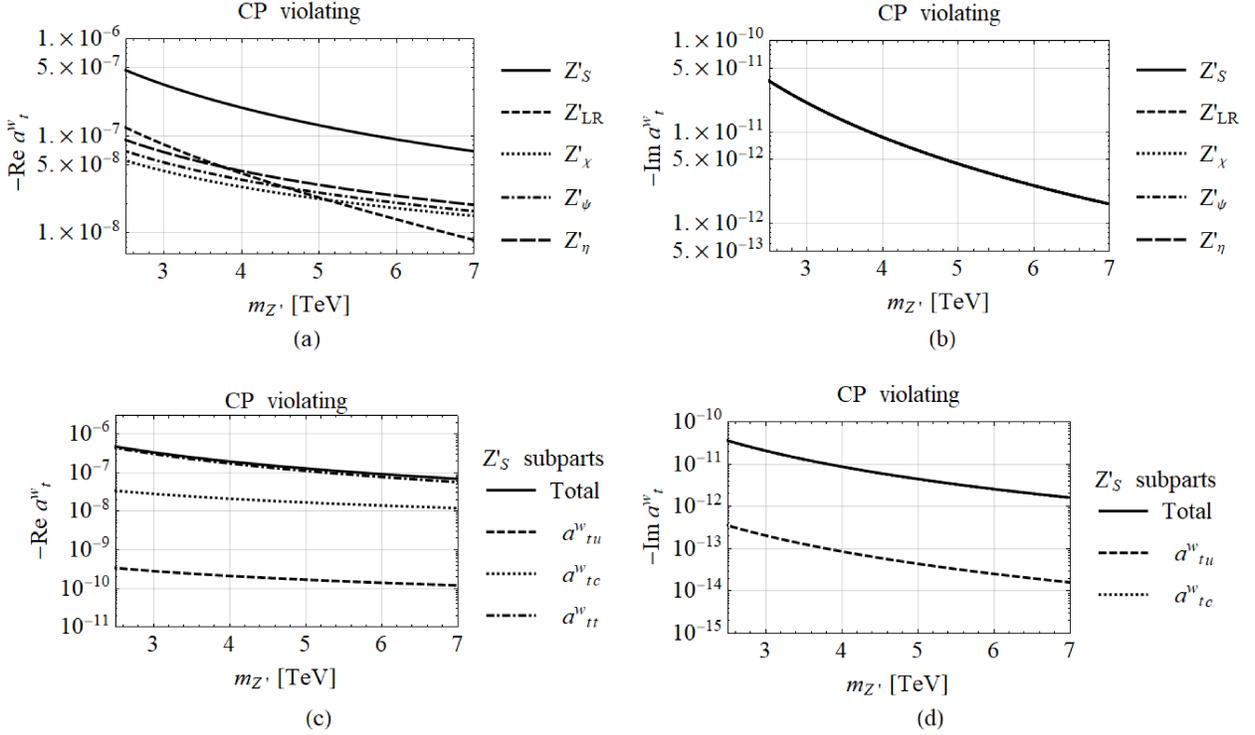}
		\caption{\small {Anomalous weak magnetic dipole moment of the top quark with CP violation, induced by a $Z^\prime$ gauge boson with flavor violation. (a) Contributions of the $Z^\prime$ gauge bosons from different models to Re ${a^w}_{t}$ and (b) the Im ${a^w}_{t}$. In  (c) and (d), it is shown the respective real and imaginary parts generated by the main contribution due the $Z^\prime_{S}$ and its subparts.}}\label{topvcpm}
	\end{figure}
\end{center}

The different values of  $-a_t^w$ as a function of the $Z^\prime$ gauge boson mass in the interval $m_{Z^\prime} =[2.5,7]$ TeV are displayed in Fig. \ref{topvcpm}.
Particularly, from Fig.  \ref{topvcpm} (a), it can be appreciated that the contributions from the  $Z^\prime$ gauge bosons to $-\mathrm{Re}\,a_t^w$, essentially show  the same behavior as  in the   CP-conserving case. In addition, we can observe that the prediction due to the $Z^\prime_S$ boson  is the most intense, being of the order of $10^{-7}$. On the other hand, the values presented by the $Z^\prime_{LR}$ boson are the least of order of $10^{-8}$.
Similarly, for the imaginary part, $\mathrm{Im}\,a_t^w$, shown in Fig. \ref{topvcpm}(b), we can appreciate that all the  $Z^\prime$ bosons share essentially the same values of $\mathrm{Im}\,a_t^w$, which go from $10^{-12}$ to $10^{-11}$.
Likewise,  Fig. \ref{topvcpm}(c) shows the  main contributing parts owing   to the  $Z^\prime_S$ boson. As we can see, the subpart $a_{tt}^w$  gives the highest value, whilst  $a_{tu}^w$ presents the  smallest one. In Fig. \ref{topvcpm} (d) it is presented the corresponding contributions  of the subparts to $-\mathrm{Im}\,a_t^w$, which are generated only by the off-diagonal $a_{tc}^w$ and $a_{tu}^w$ contributions. The contribution $a_{tt}^w$ results marginal. In both aforementioned instances, the intensities and behavior of the subparts are quite similar to that corresponding in the CP-C case.

\subsection{WEDM of the top quark}
\begin{center}
\begin{figure}[t!]
\includegraphics[scale=0.5]{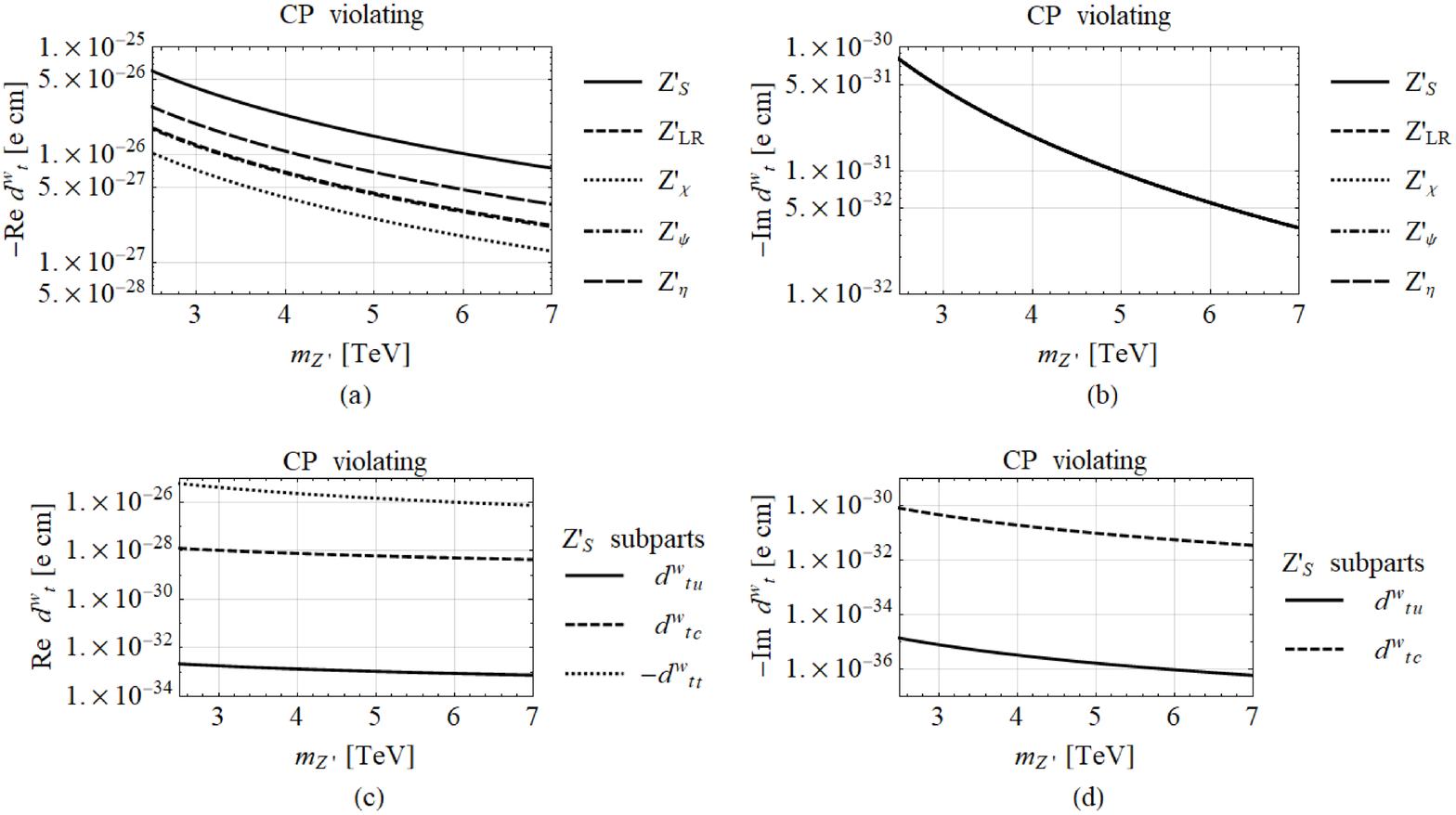}
\caption{\small {Weak electric dipole moment of the top quark with CP violation, induced by a $Z^\prime$ gauge boson with flavor violation. (a) Contributions of the $Z^\prime$  gauge bosons from different models to  $\mathrm{Re}\,{d^w}_{t}$ and (b) the  $\mathrm{Im}\,{d^w}_{t}$. In  (c) and (d) it is shown, respectively, real and imaginary parts generated by the main contribution due the $Z^\prime_{S}$ and its subparts.}}\label{topvcpe}
\end{figure}
\end{center}

In this subsection, the analysis for  WEDM of the top quark is presented, where, once again, it has been considered the  value of weak electric dipole moment in $e$-cm units. For this instance, the various contributions to $-d_t^w$ are displayed in  Fig. \ref{topvcpe} as a function of the $Z^\prime$ gauge boson mass in the interval $m_{Z^\prime} =[2.5,7]$ TeV. In Fig. \ref{topvcpe} (a) it is shown the behavior of the different contributions of $Z^\prime$ gauge bosons to  $\mathrm{Re}\,d_t^w$. As it can be observed, their absolute values decreases from $10^{-26}e$-cm to $10^{-27}e$-cm along the mass interval in discussion. Notice that the strongest prediction corresponds  to the $Z^\prime_S$ boson, while the weaker is offered by the  $Z^\prime_\chi$ gauge boson.
The behavior of $-\mathrm{Im}d_t^w$ is shown in Fig. \ref{topvcpe} (b).  We observe that the values of $-\mathrm{Im}d_t^w$ coincide for the  different $Z^\prime$ bosons, which go from $10^{-31}e$-cm to $10^{-30}e$-cm along the interval under consideration.
In Fig. \ref{topvcpe} (c), the  subparts of the main contribution owing to the $Z^\prime_S$ boson are depicted. For this case, the contribution of the subpart $-d_{tt}^w$ is the highest, in contrast, the subpart $-d_{tu}^w$ provides the smallest one.
In  Fig. \ref{topvcpe} (d), we can appreciate the behavior, as a function of $m_{Z^\prime}$,  of the  subparts that contribute to $-\mathrm{Im}d_t^w$  when the $Z^\prime_S$ is considered.  Notice that $-\mathrm{Im}d_t^w$ only receives contributions  from the off-diagonal $d_{tc}^w$ and $d_{tu}^w$ subparts, since $d_{tt}^w$ is negligible.

A final comment on  the  growing intensity of the real part  of the AWMDM and WEDM of the tau lepton, as a function of $m_{Z^\prime}$ shown in Figs. 2-4, is deserved. This is due to the also growing behavior  of the off-diagonal couplings $\Omega_{L,R f_{i}f_{j}}$ as a function of $m_{Z^\prime}$, which were studied in Refs. \cite{Aranda:2018zis,Aranda:2012qs} and \cite{Aranda:2020wfd}. The numerical analysis was done within the limits allowed by the perturbative regime.

\newpage
\section{Conclusions}

Within  the context of flavor changing neutral currents mediated by a neutral massive gauge boson, in this work we have found analytical expressions at the one-loop level for the anomalous weak magnetic and weak electric dipole moments of any charged fermion in the SM.
The theoretical framework we used, corresponds to the most general renormalizable Lagrangian that includes flavor violation mediated by a massive  gauge boson, denoted in a general manner as $Z^\prime$, which can be induced in grand unified models. Due to reasons of  interest, we restrict the numerical analysis to the case of  tau lepton and  top quark.
Because of  the CP symmetry,  we considered two cases for the AWMDM:  the CP conserving and the CP violating scenarios. Alongside  we have analyzed and determined the numerical values of the  WEDM of the tau lepton and top quark.

For the various $Z^\prime$ gauge bosons under consideration, we found that the best prediction for the $\mathrm{Re}\,a^w_{\tau}$  of the tau lepton is provided by $Z^\prime_\eta$ boson, resulting at the most of the order of  $10^{-9}$.  This value is one order of magnitude below to the $Z$ gauge boson contribution to   the SM, whose value is of the order of $10^{-8}$.
The best prediction for  $\mathrm{Im}\,a_\tau^w$ of the tau lepton comes also from the $Z^\prime_\eta$ gauge boson, resulting of the order $10^{-13}$, which is  five orders of magnitude below to the $Z$ gauge boson contribution in the SM.  The highest value for the weak electric dipole moment, is provided once again for $Z^\prime_\eta$ gauge  boson.  We found that $\mathrm{Re}\,d^w_{\tau}\sim  10^{-26}$ $e$-cm  and that $\mathrm{Im}\,d_\tau^w$ is  three orders of magnitude below.
Let us stress that our prediction for the AWMDM, due exclusively to the  sector that includes FCNC in grand unified models predicting the existence of a $Z^\prime$ gauge boson, is similar to estimates offered by extended models, such as the Unparticle Physics,  models with an  extended scalar sector or  the simplest  Little Higgs  model.

As far as the top quark is concerned,   the highest values of the AWMDM for both the real and imaginary parts, come from the $Z^\prime_S$ gauge boson. In fact, for this case the $\mathrm{Re}\, a^w_t$ is of the order of $10^{-7}$, and  $\mathrm{Im}\, a^w_t$ can reach  values of the order of  $10^{-11}$.   On  other hand, the  $\mathrm{Re}\,d^w_t$ is estimated to be between $10^{-27}$ $e$-cm and  $10^{-26}e$-cm. The respective   imaginary part is of the order of $10^{-31}$ $e$-cm.

It is interesting to note that the  numerical values of the AWMDM and WEDM  for both  the tau lepton and the top quark are not too much suppressed  with respect to that  known in the SM, when a scenario in which FCNC  is considered. According to our numerical findings, we observe that, in any of the models studied, the main contribution to the AWMDM and WEDM comes from the effect of  flavor changing.  This is because the  respective diagonal subparts  $a_{fi fj}^w$ and  $d_{fi fj}^w$, resulting from   diagonal coupling (when $fi=fj$), are  suppressed by at least two orders of magnitude   with respect to the non-diagonal ones ($fi\neq fj$), being $fi$ any of the fermions in question. This may suggests that, in the future, when the  experimental limits allow it,   measurements with more precision could evidence effects due to the presence of FCNC mediated by the  $Z^\prime$ boson to the weak dipole moments. So,  this  cannot yet be ruled out.

\section*{Appendix}
\label{sec:intro}
\noindent Form factors.
\begin{equation}
\begin{split}
F^a_V&= \frac{g}{16 \pi ^2 c_{W} e m_{Z'}^2 \left(m_{Z}^2-4 m_{f_i}^2\right)^2}
\left[m_{f_i}^4 \left(-4 m_{f_j}^2 \left(2 B_{01}+B_{03}-3 B_{02}+ C_{01} m_{Z}^2 -6 C_{01}m_{Z'}^2+1\right) \right.\right.
\\
&
\left.\left.+2 m_{Z'}^2 \left(9 B_{01}+2 B_{03}-11 B_{02}-8 C_{01} m_{Z}^2-4\right)+m_{Z}^2 (B_{02}-B_{01}+1) \right.\right.
\\
&
 \left.-14 C_{01} m_{f_j}^4-34 C_{01} m_{Z'}^4\right) -2 m_{f_i}^3 m_{f_j} \left(m_{Z'}^2 \left(6 B_{01}+4 B_{03}-10 B_{02}+8 C_{01} m_{f_j}^2-9 C_{01} m_{Z}^2\right) \right.
\\
&
\left. +2 m_{f_j}^2 (3 B_{01}-B_{02}-2 B_{03} +C_{01} (m_{f_j}^2-m_{Z}^2))-10 C_{01} m_{Z'}^4+ m_{Z}^2\right)
\\
&
+m_{f_i}^2 \left(2 m_{Z'}^4 \left(6 B_{02}+4 B_{03}-10 B_{01}-9 C_{01} m_{fj}^2+8 C_{01} m_{Z}^2\right)\right.
\\
&
 +m_{f_j}^2 m_{Z}^2 \left(2 B_{01}+B_{03}-3 B_{02}+2 C_{01} m_{f_j}^2+1\right)
\\
&
+m_{Z'}^2 \left(-m_{Z}^2 \left(9 B_{01}+ B_{03}-10 B_{02}+12 C_{01} m_{f_j}^2-2\right) \right.
\\
&
\left.+2 m_{f_j}^2 (5 B_{01}-3 B_{02}-2 B_{03})+4 C_{01} m_{Z}^4\right)
\\
&
\left. +2 m_{f_j}^4 \left(5 B_{01}-3 B_{02}-2 B_{03}+3 C_{01} m_{f_j}^2\right)+12 C_{01} m_{Z'}^6\right)
\\
&
+2 m_{f_i} m_{f_j} m_{Z}^2 \left( +m_{Z'}^2 \left(3 B_{01}+B_{03}-4 B_{02}+5 C_{01} m_{f_j}^2-2 C_{01} m_{Z}^2\right)\right.
\\
&
\left.  -m_{f_j}^2 \left(B_{03}-B_{02}+C_{01} m_{f_j}^2\right) -4 C_{01} m_{Z'}^4\right)
\\
&
+2 m_{f_i}^6 \left(B_{02}-B_{01}+C_{01} \left(5 m_{f_j}^2+m_{Z}^2+12 m_{Z'}^2\right)-2\right)
\\
&
 +2 m_{f_i}^5 m_{f_j} \left(6 B_{01}-6 B_{02}+4 C_{01} m_{f_j}^2-C_{01} m_{Z}^2 \right.
\\
&
\left.\left. -16 C_{01} m_{Z'}^2+4\right)-m_{Z}^2 (B_{01}-B_{03}) \left(m_{f_j}^4+m_{f_j}^2 m_{Z'}^2-2 m_{Z'}^4\right)-2 C_{01} m_{f_i}^8-4 C_{01} m_{f_i}^7 m_{f_j} \right].
\end{split}
\end{equation}
\begin{equation}
\begin{split}
F^a_A &= \frac{g}{16 \pi ^2 c_{W} e m_{Z'}^2 \left(m_{Z}^2-4 m_{f_i}^2\right)^2}
\left[m_{f_i}^4 \left(-4 m_{f_j}^2 \left(2 B_{01}+ B_{03}-3 B_{02}+C_{01} m_{Z}^2 \right.\right.\right.
\\
&
\left. -6 C_{01} m_{Z'}^2+1\right)+2 m_{Z'}^2 \left(9 B_{01}+2 B_{03}-11 B_{02}-8 C_{01} m_{Z}^2-4\right)
\\
&
\left. + m_{Z}^2 (B_{02}-B_{01}+1)-14 C_{01} m_{f_j}^4-34 C_{01} m_{Z'}^4\right)
\\
&
+2 m_{f_i}^3 m_{f_j} \left(m_{Z'}^2 \left(6 B_{01}+4 B_{03}-10 B_{02}+8 C_{01} m_{f_j}^2-9 C_{01} m_{Z}^2\right)\right.
\\
&
\left.+2 m_{f_j}^2 (3 B_{01}-B_{02}-2 B_{03} +C_{01} (m_{f_j}^2-m_{Z}^2) )-10 C_{01} m_{Z'}^4+m_{Z}^2\right)
\\
&
+m_{f_i}^2 \left(2 m_{Z'}^4 \left(6 B_{02}+4 B_{03}-10 B_{01}-9 C_{01} m_{f_j}^2+8 C_{01} m_{Z}^2\right)\right.
\\
&
+m_{f_j}^2 m_{Z}^2 \left(2 B_{01}+B_{03}-3 B_{02}+2 C_{01} m_{f_j}^2+1\right)
\\
&
+m_{Z'}^2 \left(-m_{Z}^2 \left(9 B_{01}+B_{03}-10 B_{02}+12 C_{01} m_{f_j}^2-2\right) \right.
\\
&
\left. +2 m_{fj}^2 (5 B_{01}-3 B_{02}-2 B_{03})+4 C_{01} m_{Z}^4\right)
\\
&
\left. +2 m_{f_j}^4 \left(5 B_{01}-3 B_{02}-2 B_{03}+3 C_{01} m_{f_j}^2\right)+12 C_{01} m_{Z'}^6\right)
\\
&
+2 m_{f_i} m_{f_j} m_{Z}^2 \left(-m_{ Z'}^2 \left(3 B_{01}+B_{03}-4 B_{02}+5 C_{01} m_{f_j}^2-2 C_{01} m_{Z}^2\right)\right.
\\
&
\left. +m_{f_j}^2 \left(B_{03}-B_{02}+C_{01} m_{f_j}^2\right)+4 C_{01} m_{Z'}^4\right)
\\
&
+2 m_{f_i}^6 \left(B_{02}-B_{01}+C_{01} \left(5 m_{f_j}^2+m_{Z}^2+12 m_{Z'}^2\right)-2\right)
\\
&
+2 m_{f_i}^5 m_{f_j} \left(6 B_{02}-6 B_{01}+C_{01} \left(-4 m_{f_j}^2+m_{Z}^2+16 m_{Z'}^2\right)-4\right)
\\
&
\left. -m_{Z}^2 (B_{01}-B_{03}) \left(m_{f_j}^4+m_{f_j}^2 m_{Z'}^2-2 m_{Z'}^4\right)-2 C_{01} m_{f_i}^8+4 C_{01} m_{f_i}^7 m_{f_j} \right].
\end{split} \label{factor fVAa}
\end{equation}
\begin{equation}
\begin{split}
F^d_{VA} & = \frac{g}{16 \pi ^2 c_{W} e m_{Z'}^2 \left(m_{Z}^2-4 m_{f_i}^2\right)^2}
\left[ m_{f_i}^4 \left(4 m_{f_j}^2 \left(3 B_{01}-2 B_{02}-B_{03} \right. \right.\right.
\\
&
\left.  -C_{01} m_{Z}^2+4 C_{01} m_{Z'}^2+1\right)+2 m_{Z'}^2 \left(9 B_{01}+2 B_{03}-11 B_{02}-8 C_{01} m_{Z}^2-4\right)
\\
&
\left. +m_{Z}^2 (B_{02}-B_{01}+1)+10 C_{01} m_{f_j}^4-34 C_{01} m_{Z'}^4\right)
\\
&
 +m_{f_i}^2 \left(2 m_{Z'}^4 \left(6 B_{02}+4 B_{03}-10 B_{01}-15 C_{01} m_{f_j}^2+8 C_{01} m_{Z}^2\right)\right.
\\
&
 +m_{Z'}^2 \left(-m_{Z}^2 \left(9 B_{01}+B_{03}-10 B_{02}+16 C_{01} m_{f_j}^2-2\right) \right.
\\
&
\left. +6 m_{f_j}^2 \left(5 B_{01}-3 B_{02}-2 B_{03}+4 C_{01} m_{f_j}^2\right)+4 C_{01} m_{Z}^4\right)+2 m_{f_j}^4 \left(3 B_{02}+2 B_{03} \right.
\\
&
\left.\left. -5 B_{01}-3 C_{01} m_{f_j}^2\right)+m_{f_j}^2 m_{Z}^2 \left(B_{03}-B_{02}+2 C_{01} m_{f_j}^2-1\right)+12 C_{01} m_{Z'}^6\right)
\\
&
+2 m_{f_i}^6 \left(B_{02}-B_{01} +C_{01} \left(-m_{f_j}^2+m_{Z}^2+12 m_{Z'}^2 \right)-2\right)
\\
&
\left. +m_{Z}^2 (B_{01}-B_{03}) \left(m_{f_j}^4-3 m_{f_j}^2 m_{Z'}^2+2 m_{Z'}^4\right)-2 C_{01} m_{f_i}^8\right].
\end{split}
\end{equation}
\begin{equation}
\begin{split}
F_{VA}^d & = \frac{i g m_{f_j}}{16 \pi ^2 c_W m_{Z'}^2 \left(4 m_{f_i}^2-m_Z^2\right)}[
m_{Z'}^2 \left(4 B_{02}-4 B_{01}+ C_{01} \left(-3 m_{f_i}^2-5 m_{f_j}^2+2 m_{Z}^2\right)\right)
\\
&
+(m_{f_i}^2-m_{f_j}^2)  (B_{02}-B_{01}+C_{01} (m_{f_i}^2- m_{f_j}^2))+4 C_{01} m_{Z'}^4 ].
\end{split}
\end{equation}
where
\begin{eqnarray*}
B_{01}&=&B_0(m_{f_i}^2,m_{Z^\prime}^2,m_{f_j}^2),\\
B_{02}&=&B_0(m_Z^2,m_{f_j}^2,m_{f_j}^2),\\
B_{03}&=&B_0(0,m_{Z^\prime}^2,m_{f_j}^2),\\
C_{01}&=&C_0(m_{f_i}^2,m_{f_i}^2,m_{Z}^2, m_{f_j}^2,m_{Z^\prime}^2, m_{f_j}^2),
\end{eqnarray*}
are the respective $B_0$ and $C_0$   Passarino-Veltman scalar functions.

\section*{ACKNOWLEDGMENTS}
This work has been partially supported by SNI-CONACYT and CIC-UMSNH. J.M. thanks to Investigadoras e Investigadores por M\'exico-CONACYT, Project 1753.


\begin{thebibliography}{99}
\bibitem{schwinger} J. Schwinger, \href{https://doi.org/10.1103/PhysRev.73.416}{\textcolor{blue}{Phys.  Rev. \textbf{73}, 416 (1948)}}..


\bibitem{kinoshita2018}T. Aoyama, T, Kinoshita, and M, Nio, \href{https://doi.org/10.1103/PhysRevD.97.036001}{\textcolor{blue}{Phys. Rev. D97, 036001 (2018)}}.

\bibitem{zeldovich1960}Ya. B. Zel'dovich, \href{http://jetp.ac.ru/cgi-bin/e/index/r/39/5/p1483?a=list}{\textcolor{blue}{Zh. Eksp. Teor. Fiz. \textbf{39}, 1483 (1960) [Sov. Phys. JETP \textbf{12}, 1030 (1960)]}}.

\bibitem{pais-primack}
A. Pais, J.R. Primack, \href{https://doi.org/10.1103/PhysRevD.8.3063}{\textcolor{blue}{Phys. Rev. D\textbf{8}, 3063 (1973)}}.


\bibitem{bernreuther} W. Bernreuther, \href{https://doi.org/10.1103/RevModPhys.63.313}{\textcolor{blue}{Rev. Mod. Phys. \textbf{63}, 313 (1991)}}.

\bibitem{cairncrossetal}W. B. Cairncross,  D. N. Gresh, M. Grau,  K. C. Cossel,  T. S. Roussy, Y. Ni,  Y. Zhou, J. Ye, and E. A. Cornell, \href{https://doi.org/10.1103/PhysRevLett.119.153001}{\textcolor{blue}{ Phys. Rev. Lett. \textbf{119}, 153001 (2017)}}.

\bibitem{opal-92} P. D. Acton \textit{et. al.}, \href{https://doi.org/10.1016/0370-2693(92)91161-2}{\textcolor{blue}{ OPAL Collab. Phys. Lett. \textbf{B281}, 405 (1992)}}.
%
\bibitem{aleph-92}  D. Buskulic \textit{et. al.}, \href{https://doi.org/10.1016/0370-2693(92)91285-H}{\textcolor{blue}{ALEPH Collab. Phys. Lett. \textbf{B297}, 459 (1992), 405 (1992)}}.
%


\bibitem{aleph-2003}
A.~Heister {\it et al.}, \href{https://doi.org/10.1140/epjc/s2003-01286-1}{\textcolor{blue}{ALEPH Collab. Eur. Phys.  J.  {\bf C30}, 291 (2003)}}.


	

\bibitem{Bernabeu:1995gs}
J.~Bernabeu, D.~Comelli, L.~Lavoura and J.~P.~Silva, \href{https://doi.org/10.1103/PhysRevD.53.5222}{\textcolor{blue}{Phys.\ Rev.\ D {\bf 53}, 5222 (1996)}}.

\bibitem{Bernabeu:1994wh}
J.~Bernabeu, G.~A.~Gonzalez-Sprinberg, M.~Tung and J.~Vidal, \href{https://doi.org/10.1016/0550-3213(94)00525-J}{\textcolor{blue}{Nucl.\ Phys.\ B {\bf 436}, 474 (1995)}}.

\bibitem{Bernabeu:1997je}
J.~Bernabeu, J.~Vidal and G.~A.~Gonzalez-Sprinberg, \href{https://doi.org/doi:10.1016/S0370-2693(97)00185-8}{\textcolor{blue}{Phys.\ Lett.\ B {\bf 397}, 255 (1997)}}.

\bibitem{Hollik:1997vb}
W.~Hollik, J.~I.~Illana, S.~Rigolin and D.~Stockinger, \href{https://doi.org/10.1016/S0370-2693(97)01259-8}{\textcolor{blue}{Phys.\ Lett.\ B {\bf 416}, 345 (1998)}}.


\bibitem{deCarlos:1997br}
B.~de Carlos and J.~M.~Moreno, \href{https://doi.org/10.1016/S0550-3213(98)00033-9}{\textcolor{blue}{Nucl.\ Phys.\ B {\bf 519}, 101 (1998)}}.

\bibitem{Hollik:1997ph}
W.~Hollik, J.~I.~Illana, S.~Rigolin and D.~Stockinger, \href{https://doi.org/10.1016/S0370-2693(98)00247-0}{\textcolor{blue}{Phys.\ Lett.\ B {\bf 425}, 322 (1998)}}.

\bibitem{GomezDumm:1999tz}
D.~Gomez-Dumm and G.~A.~Gonzalez-Sprinberg,
\href{https://doi.org/10.1007/s100529900185}{\textcolor{blue}{Eur.\ Phys.\ J.\ C {\bf 11}, 293 (1999)}}.

\bibitem{Bolanos:2013tda}
A.~Bolaños, A.~Moyotl, and G.~Tavares-Velasco, \href{https://doi.org/10.1103/PhysRevD.89.055025}{\textcolor{blue}{Phys.\ Rev.\ D {\bf 89}, 055025 (2014)}}.

\bibitem{Arroyo-Urena:2016ygo}
M.~A.~Arroyo-Ureña, G.~Hernández-Tomé, and G.~Tavares-Velasco, \href{https://doi.org/10.1140/epjc/s10052-017-4803-z}{\textcolor{blue}{Eur.\ Phys.\ J.\ C {\bf 77}, 227 (2017)}}.

\bibitem{Arroyo-Urena:2017sfb}
M.~A.~Arroyo-Ureña, G.~Tavares-Velasco and G.~Hernández-Tomé, \href{https://doi.org/10.1103/PhysRevD.97.013006}{\textcolor{blue}{Phys.\ Rev.\ D {\bf 97}, 013006 (2018)}}.


\bibitem{Arroyo-Urena:2015uoa}
M.~Arroyo-Ureña and E.~Díaz, \href{https://doi.org/10.1088/0954-3899/43/4/045002}{\textcolor{blue}{J.\ Phys.\ G {\bf 43}, 045002 (2016)}}.

\bibitem{Hollik:1998vz}
W.~Hollik, J.~I.~Illana, S.~Rigolin, C.~Schappacher and D.~Stöckinger, \href{https://doi.org/10.1016/S0550-3213(99)00201-1}{\textcolor{blue}{Nucl.\ Phys.\ B {\bf 551}, 3 (1999)}}; erratum \href{https://doi.org/10.1016/S0550-3213(99)00396-X}{\textcolor{blue}{Nucl.\ Phys.\ B {\bf 557}, 407 (1999)}}.

\bibitem{Moyotl:2012zz}
A.~Moyotl and G.~Tavares-Velasco, \href{https://doi.org/10.1103/PhysRevD.86.013014}{\textcolor{blue}{Phys.\ Rev.\ D {\bf 86}, 013014 (2012)}}.


\bibitem{buras}
A. J. Buras and J. Girrbach, \href{https://doi.org/10.1088/0034-4885/77/8/086201}{\textcolor{blue}{Rept. Prog. Phys. \textbf{77} 086201 (2014)}}.

	
\bibitem{feruglio} F. Feruglio and A. Romanino \href{https://doi.org/10.1103/RevModPhys.93.015007}{\textcolor{blue}{Rev.Mod.Phys. \textbf{93} 015007 (2021)}}.



\bibitem{SMqsectorsup} For instance, see G. Eilam, J. L. Hewett, and A. Soni, \href{https://doi.org/10.1103/PhysRevD.44.1473}{{\textcolor{blue}{Phys. Rev. D \textbf{44}, 1473 (1991)}}}; \href{https://doi.org/10.1103/PhysRevD.59.039901}{{\textcolor{blue}{\textbf{59}, 039901(E) (1998)}}}; N. G. Deshpande, B. Margolis, and H. D. Trottier, \href{https://doi.org/10.1103/PhysRevD.45.178}{{\textcolor{blue}{Phys. Rev. D \textbf{45}, 178 (1992)}}}; B. Mele, S. Petrarca, and A Soddu, \href{https://doi.org/10.1016/S0370-2693(98)00822-3}{{\textcolor{blue}{Phys. Lett. B \textbf{435}, 401 (1998)}}}; A. Cordero-Cid, J. M. Hern\'andez, G. Tavares-Velasco, and J. J. Toscano, \href{https://doi.org/10.1103/PhysRevD.73.094005}{{\textcolor{blue}{Phys. Rev. D \textbf{73}, 094005 (2006)}}}; G. Eilam, M. Frank, and I. Turan, \href{https://doi.org/10.1103/PhysRevD.73.053011}{{\textcolor{blue}{Phys. Rev. D \textbf{73}, 053011 (2006)}}}; \href{https://doi.org/10.1103/PhysRevD.74.035012}{{\textcolor{blue}{\textbf{74}, 035012 (2006)}}}.



%
\bibitem{Fukuda:1998mi}
  Y.~Fukuda {\it et al.} [Super-Kamiokande Collaboration], \href{https://doi.org/10.1103/PhysRevLett.81.1562}{{\textcolor{blue}{Phys.\ Rev.\ Lett.\  {\bf 81}, 1562 (1998)}}}.
  M.~Apollonio {\it et al.} [CHOOZ Collaboration], \href{https://doi.org/10.1140/epjc/s2002-01127-9}{{\textcolor{blue}{Eur.\ Phys.\ J.\ C {\bf 27}, 331 (2003)}}}.
  S.~N.~Ahmed {\it et al.} [SNO Collaboration], \href{https://doi.org/10.1103/PhysRevLett.92.181301}{{\textcolor{blue}{ Phys.\ Rev.\ Lett.\  {\bf 92}, 181301 (2004)}}}.
  E.~Aliu {\it et al.} [K2K Collaboration],\href{https://doi.org/10.1103/PhysRevLett.94.081802}{{\textcolor{blue}{  Phys.\ Rev.\ Lett.\  {\bf 94}, 081802 (2005)}}}.


\bibitem{cheng-li} T-P Cheng, L-F Li, Gauge Theory of Elementary Particle Physics, Clarendon Press, Oxford (1984).

\bibitem{bilenky}S. M. Bilenky and B. Pontecorvo, \href{https://doi.org/10.1016/0370-1573(78)90095-9}{{\textcolor{blue}{Phys. Rept. \textbf{41}, 225 (1978)}}}.








\bibitem{langacker1} M. Cveti\v{c}, P. Langacker, and B. Kayser, \href{https://doi.org/10.1103/PhysRevLett.68.2871}{{\textcolor{blue}{Phys. Rev. Lett. \textbf{68}, 2871 (1992)}}};
F.~Pisano and V.~Pleitez, \href{https://doi.org/10.1103/PhysRevD.46.410}{{\textcolor{blue}{Phys. Rev. D {\bf 46}, 410 (1992)}}};
P.~H.~Frampton, \href{https://doi.org/10.1103/PhysRevLett.69.2889}{{\textcolor{blue}{Phys.\ Rev.\ Lett.\  {\bf 69}, 2889 (1992)}}};
M. Cveti\v{c} and P. Langacker,
 \href{https://doi.org/10.1103/PhysRevD.54.3570}{{\textcolor{blue}{Phys. Rev. D \textbf{54}, 3570 (1996)}}}; M. Cveti\v{c}  \emph{et al.}, \href{https://doi.org/10.1103/PhysRevD.56.2861}{{\textcolor{blue}{Phys. Rev. D \textbf{56}, 2861 (1997)}}}; \href{https://doi.org/10.1103/PhysRevD.58.119905}{{\textcolor{blue}{\textbf{58}, 119905(E) (1998)}}}; M. Masip and A. Pomarol, \href{https://doi.org/10.1103/PhysRevD.60.096005}{{\textcolor{blue}{Phys. Rev. D \textbf{60}, 096005 (1999)}}};
N. Arkani-Hamed, A. G. Cohen, E. Katz, and A. E. Nelson, \href{https://doi.org/10.1088/1126-6708/2002/07/034}{{\textcolor{blue}{JHEP \textbf{07}, 034 (2002)}}}; T. Han, H. E. Logan, B. McElrath, and L.-T. Wang, \href{https://doi.org/10.1103/PhysRevD.67.095004}{{\textcolor{blue}{Phys. Rev. D \textbf{67}, 095004 (2003)}}}; C. T. Hill and E. H. Simmons, \href{https://doi.org/10.1016/S0370-1573(03)00140-6}{{\textcolor{blue}{Phys. Rept. \textbf{381}, 235 (2003)}}}; \href{https://doi.org/10.1016/j.physrep.2003.10.002}{{\textcolor{blue}{\textbf{390}, 553 (2004)}}}; J. Kang and P. Langacker, \href{https://doi.org/10.1103/PhysRevD.71.035014}{{\textcolor{blue}{Phys. Rev. D \textbf{71}, 035014 (2005)}}}; B. Fuks \emph{et al.}, \href{https://doi.org/10.1016/j.nuclphysb.2008.01.017}{{\textcolor{blue}{Nucl. Phys. \textbf{B797}, 322 (2008)}}}; J. Erler \emph{et al.}, \href{https://doi.org/10.1088/1126-6708/2009/08/017}{{\textcolor{blue}{JHEP \textbf{08}, 017 (2009)}}}; M. Goodsell \emph{et al.}, \href{https://doi.org/10.1088/1126-6708/2009/11/027}{{\textcolor{blue}{JHEP \textbf{11}, 027 (2009)}}}; P. Langacker, \href{https://doi.org/10.1063/1.3327671}{{\textcolor{blue}{AIP Conf. Proc. \textbf{1200}, 55 (2010)}}}.
\bibitem{robinett} R. W. Robinett and Jonathan L. Rosner, \href{https://doi.org/10.1103/PhysRevD.26.2396}{{\textcolor{blue}{Phys. Rev. D \textbf{26}, 2396 (1982)}}}.

\bibitem{langacker2} P. Langacker and M. Luo, \href{https://doi.org/10.1103/PhysRevD.45.278}{{\textcolor{blue}{Phys. Rev. D \textbf{45}, 278 (1992)}}}.

\bibitem{leike} A. Leike, \href{https://doi.org/10.1016/S0370-1573(98)00133-1}{{\textcolor{blue}{Phys. Rept. \textbf{317}, 143 (1999)}}}.

\bibitem{perez-soriano} M. A. P\'erez and M. A. Soriano, \href{https://doi.org/10.1103/PhysRevD.46.284}{{\textcolor{blue}{Phys. Rev. D \textbf{46}, 284 (1992)}}}.


\bibitem{ATLAS2017} M. Aaboud \textit{et al}. (ATLAS Collaboration), \href{https://doi.org/10.1016/j.physletb.2016.08.055}{{\textcolor{blue}{Phys. Lett. B \textbf{761}, 372 (2016)}}};
M. Aaboud \textit{et al}. (ATLAS Collaboration),  \href{https://doi.org/10.1007/JHEP10(2017)182}{{\textcolor{blue}{JHEP \textbf{10}, 182 (2017)}}};
M. Aaboud \textit{et al}. (ATLAS Collaboration), \href{https://doi.org/10.1016/j.physletb.2019.07.016}{{\textcolor{blue}{Phys. Lett.  B \textbf{796}, 68 (2019)}}}.

\bibitem{CMS2018} A. M. Sirunyan, A. Tumasyan, \textit{et al}. (CMS Collaboration), \href{https://doi.org/10.1007/JHEP06(2018)120}{{\textcolor{blue}{JHEP \textbf{06}, 120 (2018)}}}.


\bibitem{nowakowski} M. Nowakowski, E. A. Paschos, and J M Rodríguez,\href{https://doi.org//10.1088/0143-0807/26/4/001}{{\textcolor{blue}{Eur. J. Phys. \textbf{26}, 545 (2005)}}}.

\bibitem{Robinett:1981yz}
R.~W.~Robinett and J.~L.~Rosner, \href{https://doi.org/10.1103/PhysRevD.25.3036}{{\textcolor{blue}{Phys.\ Rev.\ D {\bf 25}, 3036 (1982)}}}; erratum \href{https://doi.org/10.1103/PhysRevD.27.679}{{\textcolor{blue}{Phys.\ Rev.\ D {\bf 27}, 679 (1983)}}}.
R.~W.~Robinett,\href{https://doi.org/10.1103/PhysRevD.26.2388}{{\textcolor{blue}{Phys.\ Rev.\ D {\bf 26}, 2388 (1982)}}}.

\bibitem{AYDE} A. Aydemir, H. Arslan, and A. K. Topaksu, \href{https://doi.org/10.1134/S1547477109040049}{{\textcolor{blue}{Phys. Part. Nucl. Lett. \textbf{6}, 304 (2009)}}}.



\bibitem{Arhrib:2006sg}
A.~Arhrib, K.~Cheung, C.~W.~Chiang, and T.~C.~Yuan, \href{https://doi.org/10.1103/PhysRevD.73.075015}{{\textcolor{blue}{Phys.\ Rev.\ D {\bf 73}, 075015 (2006)}}}.


\bibitem{feyncalc}
R. Mertig, M. B\"ohm, and  A. Denner, \href{https://doi.org/10.1016/0010-4655(91)90130-D}{{\textcolor{blue}{Comput. Phys. Commun. \textbf{64}, 345 (1991)}}}.


\bibitem{packageX}H. H. Patel, \href{https://doi.org/10.1016/j.cpc.2015.08.017}{{\textcolor{blue}{Comput. Phys. Commun. \textbf{197}, 276 (2015)}}}.

\bibitem{muon-g-2}F. Jegerlehner and A. Nyffeler, \href{https://doi.org/10.1016/j.physrep.2009.04.003}{{\textcolor{blue}{Phys. Rep. \textbf{477}, 1 (2009)}}}.



\bibitem{Aranda:2018zis}
J.~I.~Aranda, D.~Espinosa-Gómez, J.~Montaño, B.~Quezadas-Vivian, F.~Ramírez-Zavaleta, and E.~S.~Tututi, \href{https://doi.org/10.1103/PhysRevD.98.116003}{{\textcolor{blue}{Phys.\ Rev.\ D {\bf 98}, 116003 (2018)}}}.


\bibitem{Aranda:2012qs}
J.~I.~Aranda, J.~Montano, F.~Ramirez-Zavaleta, J.~J.~Toscano, and E.~S.~Tututi, \href{https://doi.org/10.1103/PhysRevD.86.035008}{{\textcolor{blue}{Phys.\ Rev.\ D {\bf 86}, 035008 (2012)}}}.




\bibitem{pdg-zyla-etal}
P.A. Zyla et. al.,  [Particle Data Group],\href{https://doi.org/10.1093/ptep/ptaa104}{{\textcolor{blue}{Prog. Theor. Exp. Phys. 2020, 083C01 (2020)}}}.


\bibitem{Aranda:2020wfd}
J.~I.~Aranda, D.~Espinosa-Gómez, J.~Montaño, F.~Ramírez-Zavaleta, and E.~S.~Tututi, \href{https://doi.org/10.1142/S0217732320501539}{{\textcolor{blue}{Mod.\ Phys.\ Lett.\ A {\bf 35},  2050153 (2020)}}}.



\bibitem{Bernreuther:1992dz}
W.~Bernreuther, T.~Schroder, and T.~N.~Pham, \href{https://doi.org/10.1016/0370-2693(92)90410-6}{{\textcolor{blue}{Phys.\ Lett.\ B {\bf 279}, 389 (1992)}}}.


\end{thebibliography}
\end{document}